\documentclass[a4paper,10pt,twoside]{cpc-hepnp}
\usepackage{CJK,upgreek,fancyhdr}
\usepackage{multicol}
\usepackage{graphicx}
\usepackage{booktabs}
\usepackage{amssymb,bm,mathrsfs,bbm,amscd}
\usepackage[tbtags]{amsmath}
\usepackage{lastpage}

\begin{document}
\begin{CJK*}{GBK}{song}

\fancyhead[c]{\small Chinese Physics C~~~Vol. xx, No. x (201x) xxxxxx}

\title{A parametrization of the cosmic-ray muon flux at sea-level\thanks{This work was partially supported by the National Natural Science Foundation of
China(10535050), the Ministry of Science and Technology of the People's
Republic of China (2006CB808102), the Research Grant Council of the Hong Kong Special Administrative Region, China (Project Nos. CUHK 1/07C and CUHK3/CRF/10), Guangdong Province and Chinese Academy of Sciences Comprehensive Strategic Cooperation Projects (2011A090100015), and the U.S. Department of Energy under Contract No. DE-AC02-05CH11231.}}

\author{%
      Mengyun Guan $^{1;1)}$\email{dreamy\_guan@ihep.ac.cn}%
\quad Ming-Chung Chu $^{2;2)}$\email{mcchu@phy.cuhk.edu.hk}%
\quad Jun Cao $^{1;3)}$\email{caoj@ihep.ac.cn}%
\quad Kam-Biu Luk $^{3;4)}$\email{k\_luk@lbl.gov}%
\quad Changgen Yang $^{1;5)}$\email{yangcg@ihep.ac.cn}%
}
\maketitle

\address{%
$^1$ Institute of High Energy Physics, Beijing 100049, China\\
$^2$ Department of Physics and Institute of Theoretical Physics, The Chinese University of Hong Kong, Shatin, Hong Kong, China\\
$^3$ Department of Physics and Lawrence Berkeley National Laboratory, University of California, Berkeley, California 94720, USA\\
}

\begin{abstract}
Based on the standard Gaisser's formula, a modified parametrization for
the sea-level cosmic-ray muon flux is introduced.
The modification is verified against experimental results.
The average vertical cosmic-ray muon intensity as a function of depth of standard rock is simulated using the modified formula as input to the MUSIC code. The
calculated muon intensities is consistent with the experimental measurements.
\end{abstract}



\footnotetext[0]{\hspace*{-3mm}\raisebox{0.3ex}{$\scriptstyle\copyright$}2013
Chinese Physical Society and the Institute of High Energy Physics
of the Chinese Academy of Sciences and the Institute
of Modern Physics of the Chinese Academy of Sciences and IOP Publishing Ltd}%

\begin{multicols}{2}

\section{Introduction}
\label{sec:intro}
In 1990, Gaisser introduced a formula for describing the cosmic-ray muon flux at sea-level \cite{Gaisser2} \cite{Gaisser},
\begin{equation}\label{eq:gaisser}
 \frac{dI_{\mu}}{dE_{\mu}}\;=\;
 0.14 \left(\frac{E_{\mu}}{GeV}\right)^{-2.7} \left[\frac{1}{1+\frac{1.1E_{\mu}\cos \theta}{115~GeV}}
  + \frac{0.054}{1+\frac{1.1E_{\mu}\cos \theta}{850~GeV}} \right ]
\end{equation}
where $I_{\mu}$ is the differential flux in units of $cm^{-2} s^{-1} sr^{-1}$, $E_{\mu}$ is
the muon energy in GeV, and $\theta$ is the zenith angle. This formula can be used for
calculating muon-induced background for various underground experiments,
such as neutrino experiments, double-$\beta$ decay experiments, dark matter search experiments, and the like.
However, this standard Gaisser's formula is only valid under the following two conditions \cite{Gaisser}:
\begin{enumerate}
  \item the curvature of the Earth can be neglected $(\theta< 70^{\circ})$.
  \item muon decay is negligible $(E_{\mu} > 100/cos \theta ~GeV)$.
\end{enumerate}

In this note, we suggest a modified parametrization that can overcome these
shortcomings.

\section{Modified parametrization}
As stated in condition 1 in the introduction, the standard Gaisser's formula
cannot be applied to all zenith angles. According to references \cite {D17}\cite{D28},
when taking the curvature of the Earth into account, the
observed zenith angle on the ground, $\theta$, and the zenith angle at the
production point of the muon in the atmosphere, $\theta^*$, are different.
Figure \ref{fig:earthcurve} illustrates the relation between these two angles.
 \begin{center}
  \includegraphics[width=7cm]{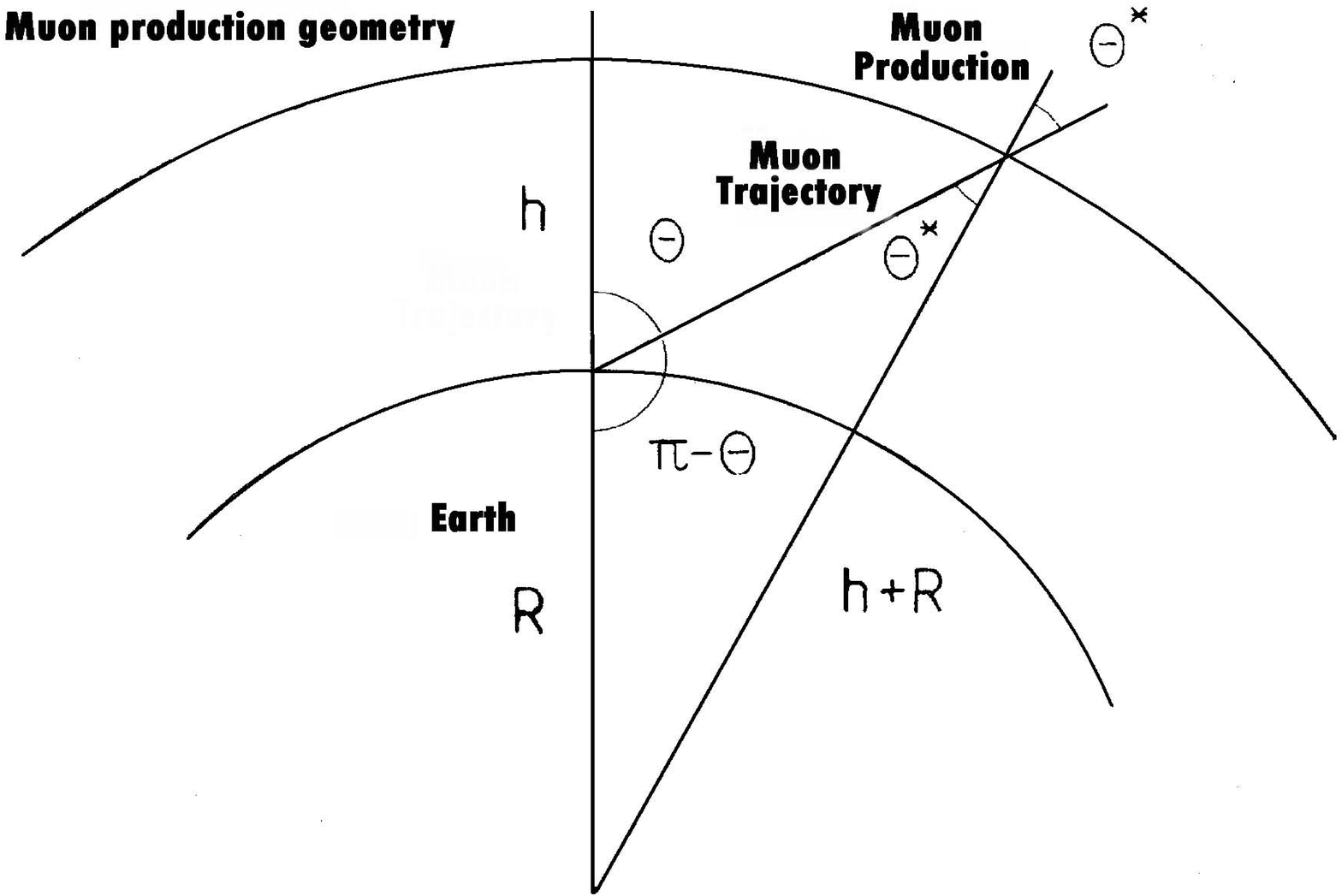}
  \figcaption{\label{fig:earthcurve}  The relation of the observed zenith angle of muons, $\theta^*$, to
   the zenith angle at the muon production point in the
   atmosphere, $\theta$. R is the radius of the Earth. Adopted from \cite {D17}\cite
{D28} }
\end{center}

Chirkin (originally tabulated in Volkova (1969)) \cite {Chirkin} pointed out that
the quantity $\cos(\theta^{\ast})$ can be related to $\cos{\theta}$ such that
\begin{equation}\label{eq:modicos}
\cos \theta^{\ast} \;=\; \sqrt{\frac{ (\cos\theta)^{2}+P_{1}^{2}+P_{2}(
\cos \theta)^{P_{3}}+P_{4}(\cos\theta)^{P_{5}}}{1+P_{1}^{2}+P_{2}+P_{4}} }\;\;
\end{equation}
where $P_1$, $P_2$, $P_3$, $P_4$, and $P_5$ are parameters given in
Table \ref{tab:par-p}.
Equation~\ref{eq:modicos} is indeed a convenient parametrization of the effect of
the Earth curvature.

\begin{center}
\tabcaption{ \label{tab:par-p} Parameters in Eq.~\ref{eq:modicos} \cite {Chirkin}.}
\footnotesize
\begin{tabular*}{80mm}{c@{\extracolsep{\fill}}ccccc}
\toprule P$_1$ & P$_2$   & P$_3$  & P$_4$ & P$_5$\\
\hline
0.102573 & -0.068287 & 0.958633 & 0.0407253 & 0.817285 \\
\bottomrule
\end{tabular*}
\vspace{0mm}
\end{center}
\vspace{0mm}

As shown in Figure \ref{fig:fit}, due to the seconds reason stated above, the standard Gaisser's formula cannot describe the experimental results well at low energies. We have modified Equation~\ref{eq:gaisser} by adding a term to the $E_{\mu}^{-2.7}$ factor.
When the muon energy increases, this term becomes negligible and the
original functional form of Gaisser is recovered.
To determine the parameters of the new term, we fit the modified equation with the zenith angle given in Equation~\ref{eq:modicos} to the world cosmic-ray muon
measurements \cite {D29} \cite {D30} \cite {D31} \cite {D32} \cite {D33}. The resulting modification is
Equation~(\ref{eq:modi}),
\begin{eqnarray}\label{eq:modi}
\frac{dI_{\mu}}{dE_{\mu}}\;=\;
 0.14 \left[ \frac{E_{\mu}}{GeV}\left(1+\frac{{3.64} GeV}{E_{\mu}(\cos \theta^{\ast})^{{1.29}}}\right) \right] ^{-2.7}  \nonumber \\ \qquad \times
 \left[\frac{1}{1+\frac{1.1E_{\mu}\cos \theta^{\ast}}{115GeV}}
  + \frac{0.054}{1+\frac{1.1E_{\mu}\cos \theta^{\ast}}{850GeV}} \right ]
\end{eqnarray}
where the parameters 3.64 and 1.29 are obtained from the fit.
The description of Equation~\ref{eq:modi} is compared to the measured
cosmic-ray muon fluxes in Figure \ref{fig:fit}. The modified parametrization
matches the experimental results fairly well for different zenith angles.

\begin{center}
  \centering  
  \includegraphics[width=8cm]{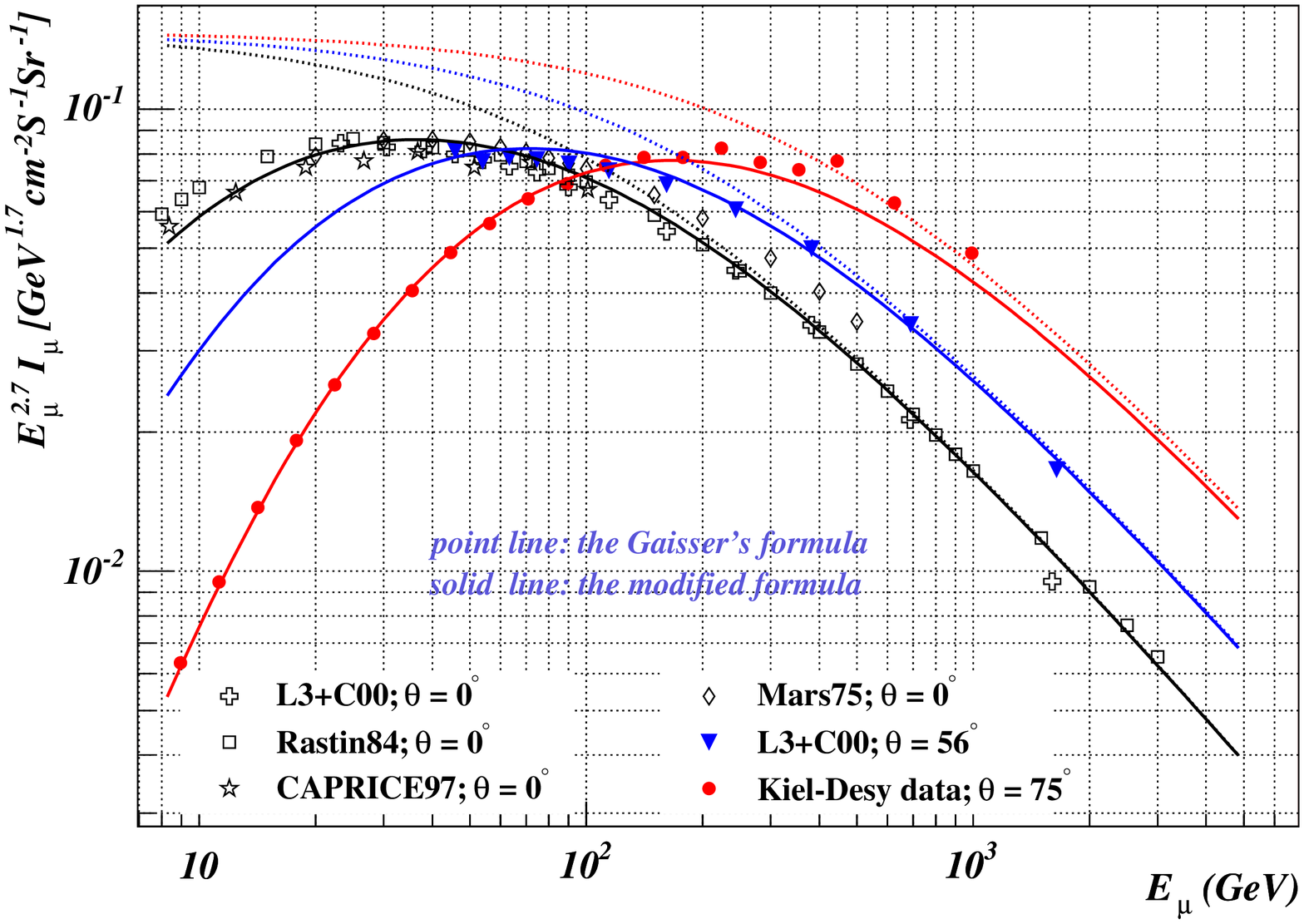}
  \figcaption{\label{fig:fit} Comparison of the modified parametrization to the
  measured cosmic-ray muon fluxes.The experimental data are from
  \cite {D29} \cite {D30} \cite {D31} \cite {D32} \cite {D33} }
\end{center}

To validate the modified parametrization, Equation~\ref{eq:modi} was
used to generate the cosmic-ray muon flux at sea level which then
served as the input to the MUSIC code \cite{MUSIC} for transporting the
simulated muons to a specific depth of standard rock.
The predicted muon flux is compared to the experimental data of vertical
muon intensity at different depths of rock overburden in
Figure \ref{fig:underground}, where the data are taken from Reference \cite {D34}.
The simulated results and experimental data agree well.
\begin{center}
  \includegraphics[width=8cm]{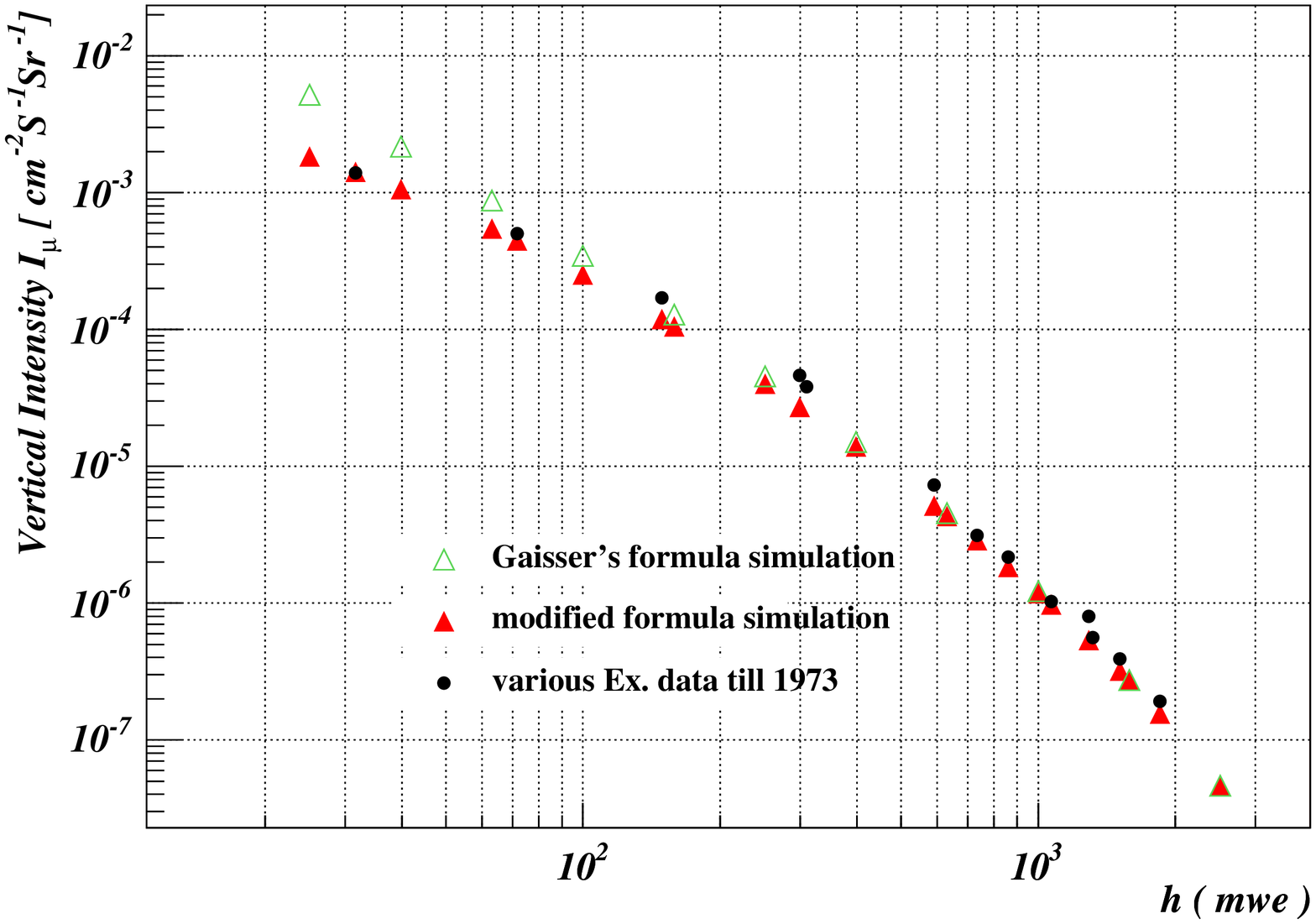}
  \figcaption{\label{fig:underground} Average vertical muon intensity versus depth of
   standard rock. Black points are experimental data from
   reference \cite {D34}. Red solid triangles stand for the simulated
   results using the modified parameterization (Equation~\ref{eq:modi}).
   Green hollow triangles are the simulated results using the standard
   Gaisser's formula (Equation~\ref{eq:gaisser}). }
\end{center}

\section{Conclusion}
We have obtained a modified Gaisser's formula that extends the range of applicability to all zenith angles and lower energies.
The new parametrization can be used conveniently and reliably for representing the cosmic-ray muon distribution at sea level for ground detectors, as well as for
underground experiments after it is coupled to a software package for transporting
the surface muons.



%
\vspace{15mm}
%

\end{multicols}

\clearpage
\end{CJK*}
\end{document}